# All-Heusler giant-magnetoresistance junctions with matched energy bands and Fermi surfaces

Zhaoqiang Bai,[1,2] Yongqing Cai,[1] Lei Shen,[1] Guchang Han,[2] and Yuanping Feng[1, †]

[1] Department of Physics, National University of Singapore, 2 Science Drive 3, Singapore 117542, Singapore

[2] Data Storage Institute, Agency for Science, Technology and Research, 5 Engineering Drive 1, Singapore 117608, Singapore

*Abstract*

We present an all-Heusler architecture which could be used as a rational design scheme for achieving high spin-filtering efficiency in the current-perpendicular-to-plane giant magnetoresistance (CPP-GMR) devices. A $Co_2MnSi/Ni_2NiSi/Co_2MnSi$ trilayer stack is chosen as the prototype of such an architecture, of which the electronic structure and magnetotransport properties are systematically investigated by first principles approaches. Almost perfectly matched energy bands and Fermi surfaces between the all-Heusler electrode-spacer pair are found, indicating large interfacial spin-asymmetry, high spin-injection efficiency, and consequently high GMR ratio. Transport calculations further confirms the superiority of the all-Heusler architecture over the conventional Heusler/transition-metal(TM) structure by comparing their transmission coefficients and interfacial resistances of parallel conduction electrons, as well as the macroscopic current-voltage (*I-V*) characteristics. We suggest future theoretical and experimental efforts in developing novel all-Heusler GMR junctions for the read heads of the next generation high-density hard disk drives (HDDs).

*Index Terms*-- **all-Heusler scheme, first-principles calculation, giant magnetoresistance, hard disk drive, read head**

[†] Author to whom correspondence should be addressed: phyfyp@nus.edu.sg

Continuous evolution of the HDD read heads, with higher sensor output, lower resistance, and higher bit resolution, is essential for further increase in the areal density in massive magnetic recording. Low-resistance magnetoresistance (MR) devices are of urgent requirement for impedance matching between read sensors and the preamplifiers, for lower electric noises, and for high frequency data transfer.[1-3] A read-sensor resistance-area product (RA) less than 0.1 $\Omega\mu m^2$ is necessary for a recording density higher than 2 Tbit/in$^2$.[2] This is a big challenge for the current magnetic tunnel junctions (MTJs) with a high impedance insulator spacer, but can be easily achieved using the current-perpendicular-to-plane giant magnetoresistance (CPP-GMR) spin valves (SVs) composed of all-metallic layers. The RA values of CPP-GMR SVs are typically below 0.05 $\Omega\mu m^2$. The drawback of the CPP-GMR SVs based on conventional ferromagnetic (FM) materials is their low signal-to-noise (SR) ratio. For example, the resistance change-area product ($\Delta$RA) of the CoFe-based CPP-GMR SV is ~1 m$\Omega$ $\mu m^2$,[4,5] which must be improved substantially. The utilization of highly spin polarized FM materials such as Co-based Heusler compounds is expected to provide large spin-dependent scatterings in the FM layers and at the interfaces between the FM and the spacer layers, thereby improving $\Delta$RA.[6] However, to the best of our knowledge, the largest room-temperature (RT) GMR ratio achieved so far is only 74.8%, which is still too low for practical application.[7] The much lower-than-expected MR value could be partially attributed to the imperfect band matching between the Heusler electrodes and TM spacer. The performance can be improved by using an all-Heusler structure with intrinsic lattice and band matching. Efforts have been made in designing and fabricating all-Heusler GMR junctions in the past decade.[8-14] Even though some design principles have been proposed and some systems have been investigated, further studies are required to understand the interface physics and identify robust device structures. In the current work, we employ a Co$_2$MnSi (CMS)/Ni$_2$NiSi (NNS)/Co$_2$MnSi trilayer GMR stack, of which the FM electrode and nonmagnetic (NM) spacer have good lattice-matching,[11] as a prototype to illustrate the physics of the electronic structure and magneto-transportation of the all-Heusler scheme and to demonstrate its advantages over the conventional Heusler/TM combination.

Figure 1 shows a model of the CMS/NNS/CMS trilayer being studied. The tetragonal supercell includes two CMS electrodes of 12 atomic layers each, and a NM NNS spacer of 1.67 nm, i.e., 13 atomic layers. For comparison, we carry out similar calculation on the CMS/Ag/CMS system with approximately the same spacer thickness (9 atomic layers of Ag). The MnSi/NiNi or MnSi/Ag interface termination was shown to be energetically stable compared to other terminations and is assumed here. The structure is optimized by relaxing the scattering region until the Hellmann-Feynman force on each atom is less than 0.01 eV/Å. First principles electronic structure calculations based on density functional theory (DFT) are performed using the Vienna Ab-initio Simulation Package (VASP),[15] whereas non-equilibrium Green's function (NEGF) method combined with DFT implemented in the Atomistix ToolKit package (ATK) is utilized for the transport calculation.[16,17] The spin-polarized generalized-gradient approximation (SGGA) proposed by Perdew et al. is employed as the exchange and correlation functional consistently throughout this work.[18] For transport calculation, the double-$\zeta$ polarized (DZP) basis set is used for the electron wave function. A cutoff energy of 150 Ry and a Monkhorst-Pack k-mesh of $8 \times 8 \times 100$ yield a good balance between computational time and accuracy in the results. It should be noted that conductance calculated using the DFT-NEGF method may deviate systematically from experimental measurement for the three-dimensional metallic multilayer structure. This, however, is not a concern here as the aim of our study is to understand the effect of the NM spacer in the FM/NM/FM trilayer structure instead of obtaining quantitative conductance of the system. Other parameters of the first-principles calculations are the same as those in Ref. 19.

The calculated majority-spin band structures of CMS and NNS and, for comparison, CMS and Ag pairs are shown in Figs. 2(a) and 2(b), respectively, whereas the corresponding Fermi surfaces are presented in Figs. 2(c)-(e), respectively. In contrast to the rather poor energy band and Fermi surface matching between CMS and Ag, the Fermi surfaces and the energy bands in the vicinity of Fermi levels of CMS and NNS almost coincide with each other. Such a good matching would ensure a smooth propagation of majority electrons across the interface, suppress the spin-flip scattering, and consequently, enhance the interfacial spin-asymmetry as well as the GMR ratio.[8, 10, 20-22]

To provide further evidence for high GMR ratio of the all-Heusler architecture and demonstrate its suitability for CPP-GMR application, we calculated the transmission spectrum and present in Fig. 3 the in-plane wave-vector $\mathbf{k}_{//}$-resolved transmission spectrum at the Fermi energy for the parallel majority spin within the CMS/NNS/CMS and CMS/Ag/CMS junctions. As can be seen in the figure, the majority transmission channel of the all-Heusler junction is largely enhanced in a vast region around the Γ point compared with that of the CMS/Ag junction. This significant enhancement of the transmittance in the parallel electrode magnetic configuration could be qualitatively attributed to the perfect Fermi surface match between the Heusler pair as shown in Fig. 2(c) and 2(d), since at the interface of perfect metallic junctions, only those propagating states with their conserved in-plane wave-vector $\mathbf{k}_{//}$ coexisting within both metals can contribute to the conduction.[23] This is also confirmed by the fact that the area of enhanced transmittance and the zone where the Fermi surfaces of CMS and NNS overlap have the same shape.

The transmission mechanism stated above can be directly visualized and verified by the macroscopic conduction behaviours, e.g., the in-plane averaged voltage drop along the transmission direction (z-axis) of the whole junction, of which the results are illustrated in Fig. 4 for the all-Heusler (red dash line) and Heusler/TM (black solid line) junctions, both in the parallel magnetization configuration. It is noted that the drop is nearly linear throughout the all-Heusler stack, which behaves as a "homogeneous" junction. However, the voltage across the CMS/Ag/CMS junction exhibits obvious heterogeneous characteristics, with a couple of step-like drops located right at the Heusler/TM interfaces, indicating comparatively heavier interfacial scattering for the majority conduction electrons. As a result, the interfacial spin-asymmetry is weakened, which, in turn, decreases ΔRA according to the Valet-Fert two-current model.[24] Moreover, this point is quantitatively confirmed by the conductance difference between the all-Heusler and Heusler/TM junctions, as listed in Tab. 1. The higher parallel interfacial conductance of the all-Heusler junction leads to a GMR ratio around 30 times larger than that of its Heusler/TM counterpart.

We also calculated the bias-dependent transmission curves of the GMR devices and obtained their macroscopic current-voltage (*I-V*) characteristics. The currents through the two junctions under parallel and antiparallel electrode magnetization configurations are calculated based on the Landauer–Buttiker formula. As can be seen in Fig. 5, the antiparallel currents of the two systems show similar trends against the bias voltage, whereas for the parallel case, there is large disparity. The parallel current of the all-Heusler junction is, unsurprisingly, much larger that of the Heusler/TM one, which can be attributed to its aforementioned smaller interfacial resistance. This explicitly confirms that the all-Heusler interfaces have much improved spin-valve performance compared with the widely-used Heusler/TM interfacial structure.

In summary, we present an all-Heusler architecture for the rational design of high-efficiency GMR junctions for the HDD read heads. The intrinsic advantages of the all-Heusler structure, i.e., the almost perfectly matched energy bands and Fermi surfaces between the electrodes and spacer, are expected to lead to much higher spin-injection efficiency at the FM/NM interface, and hence a much larger $\Delta RA$ compared with its Heusler/TM counterparts. It should be noted that we discuss the all-Heusler CMS/NNS junction only as a representative example of an optimum design scheme superior to the state of art. Further theoretical and experimental efforts, following the all-Heusler scheme, are strongly recommended in designing and fabricating well-crystallized NM Heusler compounds as the tag spacer materials which can well match the Co-based full-Heusler electrodes.

**Figure caption**

Fig.1 A schematic device model of the $Co_2MnSi/Ni_2NiSi/Co_2MnSi$ giant magnetoresistance junction.

Fig.2 The majority spin band structures of (a) CMS (black solid line) vs. NNS (red dash line) and (b) CMS (black solid line) vs. Ag (red dash line). The Fermi surfaces in the first Brillouin zones corresponding to the tetragonal unit cells of (c) $L2_1$-CMS, (d) $L2_1$-NNS, and (e) fcc-Ag plotted by XCRYSDEN[24].

Fig.3 In-plane wave vector $\mathbf{k}_{//}=(k_a, k_b)$ dependence of the majority spin transmittance at the Fermi level for (a) CMS/NNS/CMS and (b) CMS/Ag/CMS GMR junction in the parallel magnetization configuration.

Fig.4 In-plane averaged voltage drop along the transmission direction (z-axis) across the all-Heusler (red dash line) and Heusler/TM (black solid line) junctions. The vertical black dash lines show the position of the FM/NM interfaces.

Fig.5 Calculated *I-V* curves of the $Co_2MnSi/Ni_2NiSi/Co_2MnSi$ and $Co_2MnSi/Ag/Co_2MnSi$ GMR junctions.

**Table**

Tab. 1 The conductance (in Siemens) of the CMS/NNS/CMS and CMS/Ag/CMS GMR junctions in parallel and antiparallel magnetization configurations.

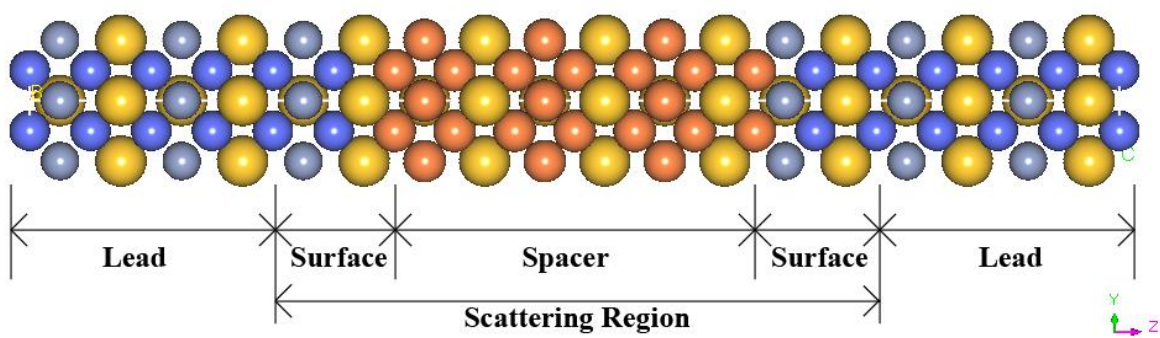

**Figure 1**

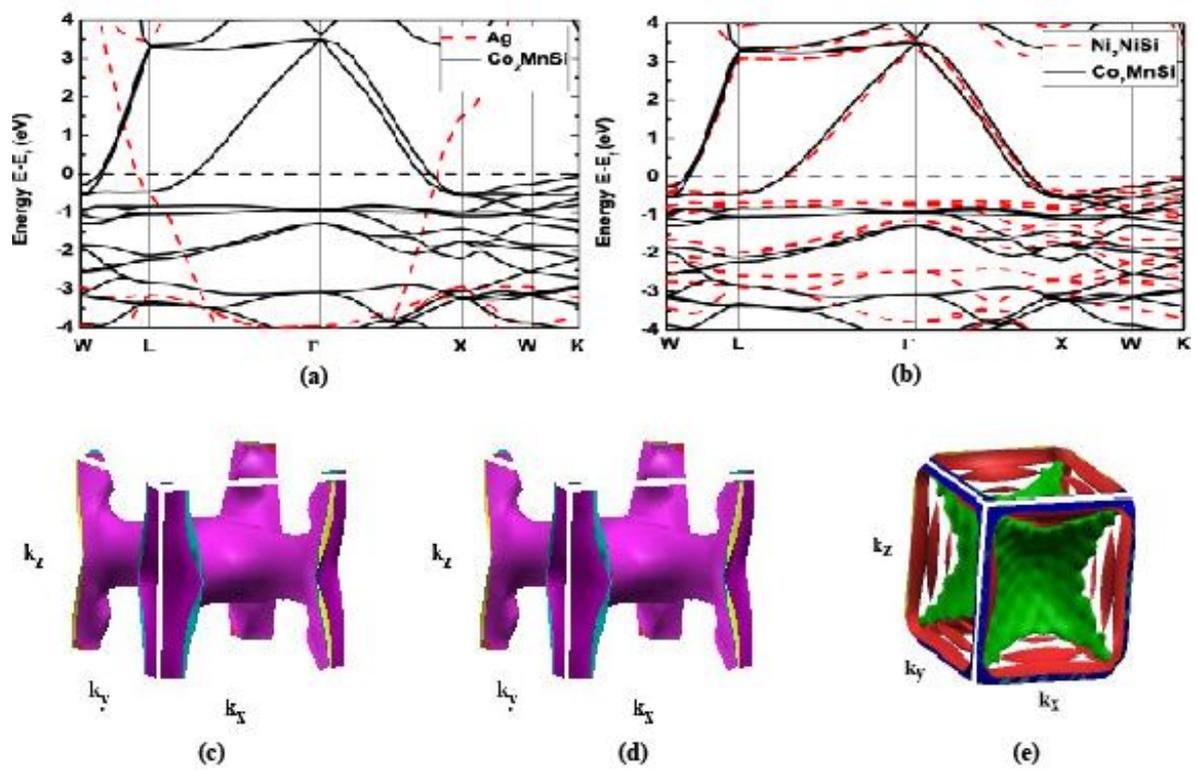

**Figure 2**

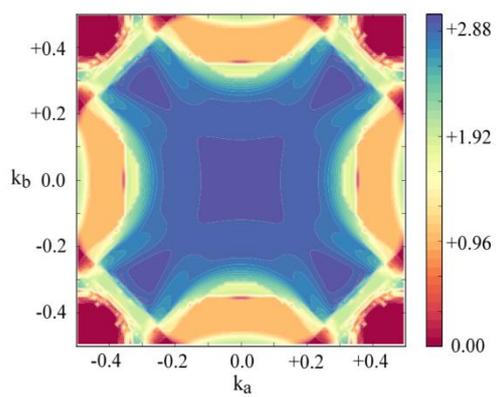 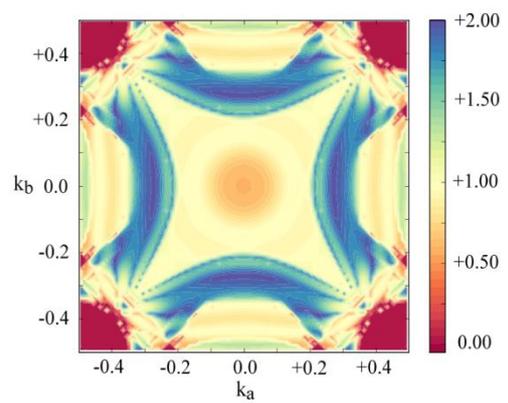

**(a)** **(b)**

**Figure 3**

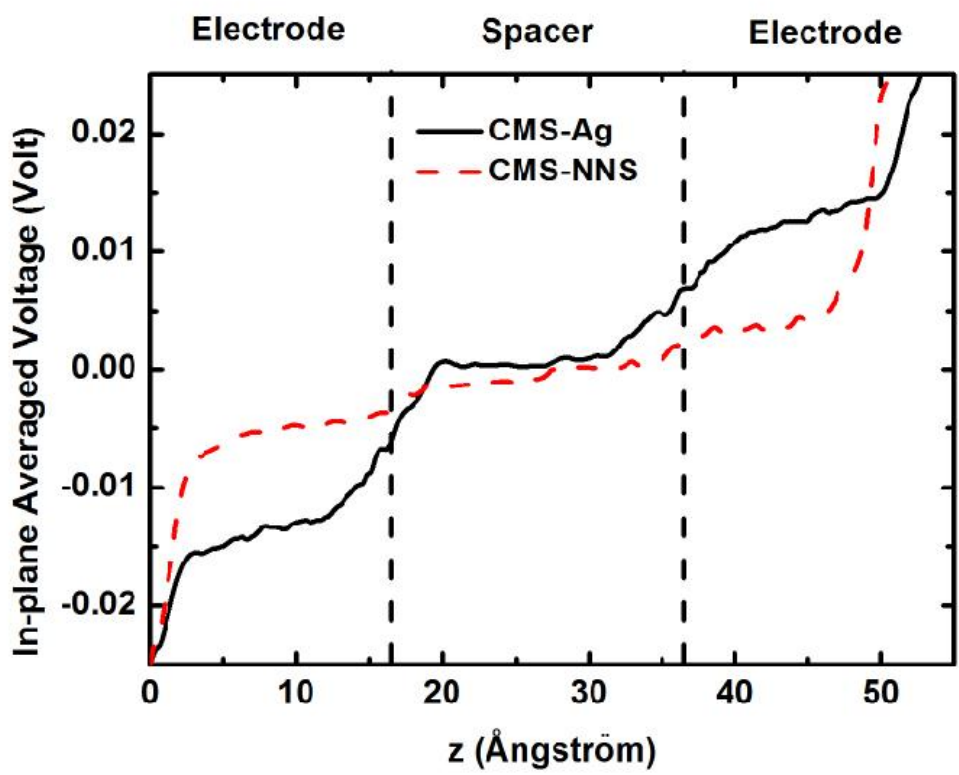

**Figure 4**

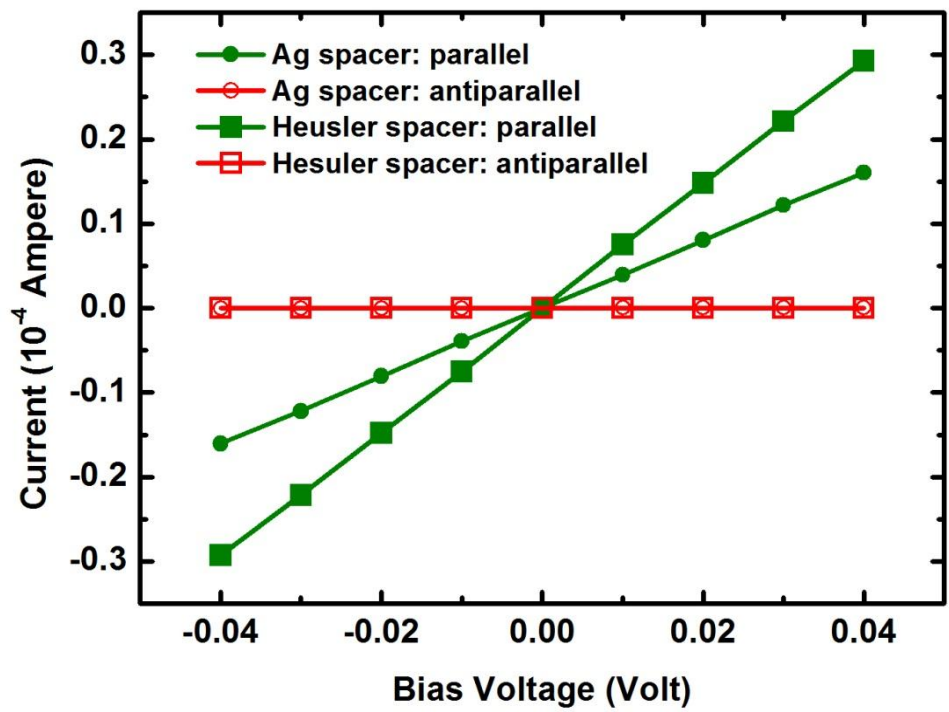

**Figure 5**

| Structure | $G_P$ | $G_{AP}$ |
|---|---|---|
| **CMS/NNS/CMS** | $6.71 \times 10^{-4}$ | $1.48 \times 10^{-14}$ |
| **CMS/Ag/CMS** | $2.34 \times 10^{-5}$ | $1.39 \times 10^{-14}$ |

**Table 1**